# Treatment of the singularities of the Sternheimer function for a layered electron gas

*D. E. Beck*
*Department of Physics and Astronomy, University of Hawaii-Manoa, HI 96822, U. S. A.*

**Abstract**
Isolating the singularity in the Green's function solution of the inhomogeneous, differential equation for the Sternheimer function of a layered electron gas, permits the construction of an approximate solution of the Sternheimer function which includes the exact contributions from its singularities.

# I. Introduction

The Modified Sternheimer procedure [1, 2] is employed in many computation of the density response in the time-dependent density functional theory (TDDFT) [3, 4]. The susceptibility of the induced density in the theory has poles [3] and these poles contribute singularities to the Sternheimer function. This function satisfies an inhomogeneous Schrodinger equation and, in Sec. III, we proceed by constructing the Green's function solution of this equation. Our construction of the Green's function isolates its singularity and permits a useful approximation of its angular integral. Utilizing this approximation we obtain an expression for the angular integral of the Sternheimer function which includes the contributions from the singularities (15 and 40). In Sec. IV, we demonstrate that our approximation includes the exact contributions from the singularities of the Sternheimer function, (43 and 44). In Sec. II, we consider a model for a layered electron gas [4, 5], and our approximation for the Sternheimer function is developed for this system.

# II. Susceptibility and the Sternheimer Function

The induced density in the TDDFT [4] is expressed as,

$$\rho(\mathbf{r};\omega) = \int d\mathbf{r}'\chi(\mathbf{r},\mathbf{r}';\omega)v_{scf}(\mathbf{r}';\omega), \tag{1}$$

and for a layered electron system with the electrons confined in uniform, parallel layers the susceptibility is given by [4,5],

$$\chi(\mathbf{r},\mathbf{r}';\omega) = \Omega^{-2}\sum_{\mathbf{k},\mathbf{k}'}\frac{f(\mathbf{k})-f(\mathbf{k}')}{E_{\mathbf{k}}-E_{\mathbf{k}'}+\hbar\omega}e^{i(\mathbf{K}'-\mathbf{K})\cdot(\mathbf{R}-\mathbf{R}')}\psi^*_{k_z n}(z)\,\psi_{k'_z n'}(z)\,\psi^*_{k'_z n'}(z')\,\psi_{k_z n}(z'). \tag{2}$$

The energy is $E_{\mathbf{k}} = \xi_{k_z n} + \frac{\hbar^2}{2m}|\mathbf{K}|^2$, and $f(\mathbf{k}) = f(E_{\mathbf{k}})$ is the zero-temperature Fermi distribution function which restricts the band energies to be less than the Fermi energy, $\varepsilon_{k_z n} \le E_f$. The volume is $\Omega = L^2(Nd)$ and we employing the Bloch ansatz,

$$\int d\mathbf{R}e^{-i\mathbf{Q}.\mathbf{R}}\rho(\mathbf{r};\omega) = e^{iq_z dl}\rho_{q_z}(z;Q,\omega).$$

The self-consistent-potential is the time-dependent-local density response induced by the perturbing potential,

$$\int d\mathbf{R}e^{-i\mathbf{Q}.\mathbf{R}}v_{scf}(\mathbf{r};\omega) = e^{iq_z dl}v_{q_z}(z;Q,\omega).$$

The wave functions are the eigenfunctions of the Schrodinger equation,

$$\mathrm{L}_0\ \psi_{k_z n}(z) = [-\tfrac{\hbar^2}{2m}\partial_z^2 + v_{eff}(z)]\psi_{k_z n}(z) = \xi_{k_z n}\psi_{k_z n}(z)\ . \tag{3}$$

The effective potential is obtained using the Kohn-Sham procedure, and it depends on the computed density. The periodic, boundary conditions for the band states are,

$$\psi_{k_z n}(d/2) - e^{ip_z d}\,\psi_{k_z n}(-d/2) = 0\,, \tag{4a}$$

and,

$$\partial_z\psi_{k_z n}(d/2) - e^{ip_z d}\partial_z\psi_{k_z n}(-d/2) = 0\,, \tag{4b}$$

and the states are normalized,

$$\tfrac{1}{d}\int_{-d/2}^{d/2} dz\left|\psi_{k_z n}(z)\right|^2 = 1\,.$$

Performing the integrals over parallel wave-numbers, results in a z-dependent equation for the induced density response,

$$\rho_{q_z}(z;Q\omega) = \tfrac{1}{d}\int_{-d/2}^{d/2} dz'[\chi_{q_z}^{+}(z,z';Q\omega) + \chi_{q_z}^{-}(z,z';Q\omega)]v_{q_z}(z';Q\omega)\,, \tag{5}$$

with the linear susceptibility given by,

$$\begin{aligned}\chi_{q_z}^{\pm}(z,z';Q\omega) = \tfrac{1}{d^2}\sum_{k_z n}\frac{K_{\max}^2}{2\pi}\sum_{k_z' n'}\psi_{k_z n}^{*}(z)\,\psi_{k_z' n'}(z)\psi_{k_z' n'}^{*}(z')\,\psi_{k_z n}(z')\delta(k_z - k_z' + q_z)\\ \times\tfrac{1}{2\pi}\int_{-\pi}^{\pi} d\vartheta\frac{\sin^2\vartheta}{\xi_{k_z n} - \xi_{k_z' n'} - \tfrac{\hbar^2}{2m}(2K_{\max}Q\cos\vartheta + Q^2) \pm \hbar(\omega + i\varepsilon)}\end{aligned}\ . \tag{6}$$

The integrations over the parallel wave-numbers and the Fermi distribution function provide the cutoff, $K_{\max}^2 = \tfrac{2m}{\hbar^2}(E_f - \varepsilon_{k_z n}) \geq 0$ .

The angular integral can be analytically evaluated [6],

$$\begin{aligned}\tfrac{1}{2\pi}\int_{-\pi}^{\pi} d\vartheta\sin^2\vartheta\frac{\delta(k_z + q_z - k_z')}{\xi_{k_z n} - \xi_{k_z' n'} - \tfrac{\hbar^2}{2m}[2K_{\max}Q\cos\vartheta + Q^2] \pm \hbar(\omega + i\varepsilon)} =\\ = \frac{1}{\tfrac{\hbar^2}{m}K_{\max}Q}\tfrac{1}{2\pi}\int_{-\pi}^{\pi} d\vartheta\sin^2\vartheta\frac{1}{\cos\vartheta_{k_z+q_z n'}^{\pm} - \cos\vartheta} = -\frac{S(-\cos\vartheta_{k_z+q_z n'}^{\pm})}{\tfrac{\hbar^2}{m}K_{\max}Q}\end{aligned}, \tag{7}$$

where

$$\cos\vartheta_{k_z+q_z n'}^{\pm} = [\xi_{k_z n} - \tfrac{\hbar^2}{2m}Q^2 \pm \hbar(\omega + i\varepsilon) - \xi_{k_z+q_z n'}]/\tfrac{\hbar^2}{m}K_{\max}Q\,.$$

Then the linear susceptibility is given by,

$$\begin{aligned}\chi_{q_z}^{\pm}(z,z';Q\omega) = \tfrac{1}{d}\sum_{k_z n}\frac{K_{\max}^2}{2\pi}\psi_{k_z n}^{*}(z)\,\psi_{k_z n}(z')\\ \times\frac{1}{\tfrac{\hbar^2}{m}K_{\max}Q}\tfrac{1}{d}\sum_{n'}\psi_{k_z+q_z n'}(z)\,\psi_{k_z+q_z n'}^{*}(z')S(\cos\vartheta_{k_z+q_z n'}^{\pm})\end{aligned}\ . \tag{8}$$

The difficulty encountered in utilizing this result is that, for the real part of the integral, the sum over $n'$ is unrestricted and the contributions from all the unoccupied states must be computed. This problem is circumvented by employing a Modified Sternheimer procedure [1] to provide an evaluation of the summation over $n'$. The Sternheimer function is defined as,

$$\zeta_{p_zn}(z;E)=\sum_{n'}\frac{\psi_{p_zn'}(z)}{\xi_{p_zn'}-E}\Delta^*_{p_zn',n}\,, \tag{9a}$$

where,

$$\Delta^*_{p_zn',n}=\tfrac{1}{d}\int_{-d/2}^{d/2}dz'\psi^*_{p_zn'}(z')\psi_{k_zn}(z')v_{q_z}(z')\delta(k_z+q_z-p_z)\,. \tag{9b}$$

The singularities of the Sternheimer function occur at the band energies, $E=\xi_{p_zn'}$. Utilizing this function, the induced density is determined by,

$$\rho^{\pm}_{q_z}(z;Q,\omega)=-\tfrac{1}{d}\sum_{k_zn}\frac{K^2_{\max}}{2\pi}\psi^*_{k_zn}(z)\tfrac{1}{2\pi}\int_{-\pi}^{\pi}d\vartheta\sin^2\vartheta\zeta_{p_zn}(z;E^{\pm}_{n,\cos\vartheta})\,, \tag{10a}$$

where the energy is,

$$E^{\pm}_{n,\cos\vartheta}=\xi_{k_zn}-\tfrac{\hbar^2}{2m}(2K_{\max}Q\cos\vartheta+Q^2)\pm\hbar(\omega+i\varepsilon)\,. \tag{10b}$$

Employing the Schrodinger operator, one obtains an inhomogeneous, differential equation for the Sternheimer function,

$$\mathrm{L}_E\,\zeta_{p_zn}(z;E)=[-\tfrac{\hbar^2}{2m}\partial_z^2+v_{eff}(z)-E]\zeta_{p_zn}(z;E)=\psi_{k_zn}(z)v_{q_z}(z)\,. \tag{11}$$

Here $p_z=k_z+q_z$, $\psi_{k_zn}(z)$ is an occupied band state, and we have employed the completeness of the band states,

$$\tfrac{1}{d}\sum_{n'}\psi_{k'_zn'}(z)\psi^*_{k'_zn'}(z')=\delta(z-z')\,.$$

The periodic, boundary conditions are the same as those for the band states,

$$\zeta_{p_zn}(d/2;E)-e^{ip_zd}\zeta_{p_zn}(-d/2;E)=0\,, \tag{12a}$$

and,

$$\partial_z\zeta_{p_zn}(d/2;E)-e^{ip_zd}\partial_z\zeta_{p_zn}(-d/2;E)=0\,. \tag{12b}$$

In the next section we construct a Green's function for this Schrodinger operator which provides a solution for the Sternheimer function,

$$\zeta_{p_zn}(z,E)=\tfrac{2m}{\hbar^2}\int_{-d/2}^{d/2}dz'G_{p_zE}(z,z')\psi_{k_zn}(z')v_{q_z}(z')\,. \tag{13}$$

The Green's function with its boundary conditions is unique. However, the choice of the functions used in the construction of the Green's function is not unique [7]. Our choice of these functions and their arrangement permits us to isolate the singularities of the Green's function, (19 and 20). The wave functions utilized in our construction are solutions of,

$$L_E\tilde{\psi}_{p_zE}(z)=0\,. \tag{14a}$$

They satisfy the boundary condition,

$$\tilde{\psi}_{p_zE}(d/2)-e^{ip_zd}\tilde{\psi}_{p_zE}(-d/2)=0\,, \tag{14b}$$

and the non-periodic condition,

$$\partial_z\tilde{\psi}_{p_zE}(d/2)-e^{ip_zd}\partial_z\tilde{\psi}_{p_zE}(-d/2)=Z_{p_zE}=\gamma_{p_zE}/\tilde{\psi}^*_{p_zE}(d/2)\,, \tag{14c}$$

in the interval, $-d/2\le z\le d/2$. Their normalization is,

$$N^2_{p_zE}=\tfrac{1}{d}\int_{-d/2}^{d/2}dz\left|\tilde{\psi}_{p_zE}(z)\right|^2$$

The construction permits an approximation for the angular integral of the Green's function [See (35), (38) and (39)] and the Sternheimer function,

$$\overline{\zeta}_{p_z n}(z,E_n^{\pm}) = \tfrac{1}{2\pi}\int_{-\pi}^{\pi} d\vartheta \sin^2\vartheta\, \zeta_{p_z n}(z,E_{n,\cos\vartheta}^{\pm})$$

$$\approx g_{p_z E_n^{\pm}} \tfrac{2m}{\hbar^2}\int_{-d/2}^{d/2} dz' G_{p_z E_n^{\pm}}(z,z')\,\psi_{k_z n}(z') v_{q_z}(z'). \tag{15a}$$

$$= g_{p_z E_n^{\pm}} \zeta_{p_z n}(z,E_n^{\pm})$$

The energy is,

$$E_n^{\pm} = \xi_{k_z n} - \tfrac{\hbar^2}{2m} Q^2 \pm \hbar(\omega + i\varepsilon), \tag{15b}$$

the coefficient is,

$$g_{p_z E_n} = \frac{\gamma_{p_z E_n}}{dN_{p_z E_n}^2} \frac{S(\alpha_{p_z E_n} \mp i\,\mathrm{sgn}\,\varepsilon)}{2K_{\max} Q}, \tag{15c}$$

and the scaling parameter is,

$$\alpha_{p_z E} = \tilde{\psi}_{p_z E}^{*}(d/2) Z_{p_z E} / (2K_{\max} Q dN_{p_z E}^2). \tag{16}$$

Hence, we obtain our approximation for the induced density,

$$\rho_{q_z}^{\pm}(z;Q,\omega) \approx -\tfrac{1}{d}\sum_{k_z n} \frac{K_{\max}^2}{2\pi} \psi_{k_z n}^{*}(z) g_{p_z E_n^{\pm}} \zeta_{p_z n}(z;E_n^{\pm}). \tag{17}$$

The Sternheimer function, (9), is a solution of (11) and singular. However, we show in Sec. IV, (40), that the results of the integration over the singularities has been incorporated into the analytic evaluation of the integral, $S(\alpha_{p_z E_n} \mp i\,\mathrm{sgn}\,\varepsilon)$.

## III. Green's Function and the Angular Integral

When the energy lies within a band, $E = \xi_{p_z n'}$, the band state, $\psi_{p_z n'}(z)$, is a solution of the homogenous equation which satisfies the periodic boundary conditions. Therefore, the functions we use to construct the Green's function are required to satisfy an additional condition in order to have a unique solution for the non-homogenous equation [7]. In order to establish this condition, we evaluate,

$$\Delta_{p_z E,n}^{*} = \tfrac{1}{d}\int_{-d/2}^{d/2} dz\, \tilde{\psi}_{p_z E}^{*}(z)\psi_{k_z n}(z) v_{q_z}(z)\delta(k_z + q_z - p_z)$$

$$= \tfrac{1}{d}\int_{-d/2}^{d/2} dz\, \tilde{\psi}_{p_z}^{*}(z)\, \mathrm{L}_E\, \zeta_{p_z n}(z,E)$$

$$= -\tfrac{\hbar^2}{2md}\Big(\tilde{\psi}_{p_z E}^{*}(z)\partial_z \zeta_{p_z n}(z,E) - \zeta_{p_z n}(z,E)\partial_z \tilde{\psi}_{p_z E}^{*}(z)\Big]_{-d/2}^{d/2}. \tag{18}$$

$$+ \tfrac{1}{d}\int_{-d/2}^{d/2} dz\, \zeta_{p_z n}(z,E)\, \mathrm{L}_E\, \tilde{\psi}_{p_z E}^{*}(z)$$

$$= \tfrac{\hbar^2}{2md}\Big(\zeta_{p_z n}(z,E)\partial_z \tilde{\psi}_{p_z E}^{*}(z)\Big]_{-d/2}^{d/2} = \tfrac{\hbar^2}{2md} Z_{p_z E}^{*} \zeta_{p_z n}(d/2, E)$$

Since $\lim_{E\to\xi_{p_z n'}} Z_{p_z E}^{*} = 0$ when the energy coincides with that of a band state, the required condition that $\Delta_{p_z \xi_{p_z n'},n}^{*} = 0$ for a unique solution is satisfied by our wave functions (14).

We factor the Green's function as,

$$G_{p_z E}(z,z') = G_{p_z E}^{(1)}(z,z') + \gamma_{p_z E}^{-1} \tilde{\psi}_{p_z E}(z) \tilde{\psi}_{p_z E}^{*}(z'), \tag{19}$$

in order to isolate its singularity. Here, $G_E^{(1)}(z,z')$ is constructed so that it vanishes at the boundaries,

$$G_E^{(1)}(d/2, z') = G_E^{(1)}(-d/2, z') = 0\,, \tag{20a}$$
$$G_E^{(1)}(z, d/2) = G_E^{(1)}(z, -d/2) = 0\,, \tag{20b}$$
and [8],
$$\mathrm{L}_E\, G_E^{(0)}(z,z') = \mathrm{L}_E\, G_E^{(1)}(z,z') = \tfrac{\hbar^2}{2m}\,\delta(z-z')\,. \tag{21}$$
For this choice of $G_E^{(1)}(z,z')$, the boundary conditions on its derivatives are,
$$\partial_z G_E^{(1)}(z,z')\Big|_{z=d/2} - e^{ipd}\,\partial_z G_E^{(1)}(z,z')\Big|_{z=-d/2} = -\tilde{\psi}^*_{p_zE}(z')\,/\,\tilde{\psi}^*_{p_zE}(d/2)\,, \tag{22a}$$
and,
$$\partial_{z'} G_E^{(1)}(z,z')\Big|_{z'=d/2} - e^{-ipd}\,\partial_{z'} G_E^{(1)}(z,z')\Big|_{z'=-d/2} = -\tilde{\psi}_{p_zE}(z)\,/\,\tilde{\psi}_{p_zE}(d/2)\,. \tag{22b}$$
Hence, the boundary condition on the derivative of the Green's function requires
$$\gamma_{p_zE} = [\tilde{\psi}^*_{p_zE}(d/2)\partial_z\tilde{\psi}_{p_zE}(d/2) - \tilde{\psi}^*_{p_zE}(-d/2)\partial_z\tilde{\psi}_{p_zE}(-d/2)] = Z_{p_zE}\,\tilde{\psi}^*_{p_zE}(d/2)\,. \tag{23a}$$
and,
$$\gamma_{p_zE} = [\tilde{\psi}_{p_zE}(d/2)\partial_z\tilde{\psi}^*_{p_zE}(d/2) - \tilde{\psi}_{p_zE}(-d/2)\partial_z\tilde{\psi}^*_{p_zE}(-d/2)] = Z^*_{p_zE}\,\tilde{\psi}_{p_zE}(d/2)\,. \tag{23b}$$
The singularities of the Green's function are isolated in this coefficient, since the singularities are located at the band energies, and $\lim_{E\to\xi_{p_zn'}} Z^*_{p_zE} = 0$. Employing the Wronskian,
$$W = \tilde{\psi}_{p_zE}(z)\partial_z\tilde{\psi}^*_{p_zE}(z) - \tilde{\psi}^*_{p_zE}(z)\partial_z\tilde{\psi}_{p_zE}(z) = \text{constant}\,,$$
for the solutions to the homogenous equation, we can equate the two expressions, and we obtain a more insightful expression for the evaluation of our approximation,
$$\gamma_{p_zE} = \tfrac{1}{2}\partial_z\left|\tilde{\psi}_{p_zE}(z)\right|^2\Big|_{-d/2}^{d/2}\,. \tag{24}$$
We simplify our computation by introducing approximations for the angular integral of the Green's function. Since $G_E^{(1)}(z,z')$ has no singularities as a function of the energy, we approximate its angular integral by,
$$\tfrac{1}{2\pi}\int_{-\pi}^{\pi} d\vartheta \sin^2\vartheta\, G^{(1)}_{E_{n,\cos\vartheta}}(z,z') \approx g_{p_zE_{n,\cos\vartheta}}\, G^{(1)}_{E_{n,\cos\vartheta}}(z,z')\Big|_{\cos\vartheta_0}\,. \tag{25}$$
The remaining term has a singularity when the energy coincides with that of a band state, and $\gamma_{p_zE}\big|_{E=\xi_{p_zn'}} = 0$. In order to approximate the contribution from this integral and provide for the singularity, we consider,
$$\left.\frac{\tilde{\psi}_{p_zE}(z)\tilde{\psi}^*_{p_zE}(z')}{\tfrac{1}{2}\left(\partial_z\left|\tilde{\psi}_{p_zE}(z)\right|^2\Big|_{-d/2}^{d/2}\right)}\right|_{E=E_{n,\cos\vartheta}} \approx \left.\frac{\tilde{\psi}_{p_zE}(z)\tilde{\psi}^*_{p_zE}(z')}{\tfrac{1}{2}\left|\tilde{\psi}_{p_zE}(d/2)\right|^2}\right|_{E=E_{n,\cos\vartheta_0}} \frac{1}{\partial_z \ln\left|\tilde{\psi}_{p_zE_{n,\cos\vartheta}}(z)\right|^2\Big|_{-d/2}^{d/2}}\,. \tag{26}$$
The leading factor in this approximation has no explicit dependence on the singularities and we only want to integrate over the logarithmic term. Hence, we need to determine its energy dependence,
$$\partial_E\left(\partial_z \ln\left|\tilde{\psi}_{p_zE}(z)\right|^2\Big|_{-d/2}^{d/2}\right) = \partial_E\,\frac{Z_{p_zE}\,\tilde{\psi}^*_{p_zE}(d/2)}{\tfrac{1}{2}\left|\tilde{\psi}_{p_zE}(d/2)\right|^2} = 2\,\frac{[\partial_E Z_{p_zE} - Z_{p_zE}\partial_E \ln\tilde{\psi}_{p_zE}(d/2)]}{\tilde{\psi}_{p_zE}(d/2)}\,. \tag{27}$$

The differential equation for the wave function provides an expression for the energy dependence of the wave function,

$$\partial_E L_E \tilde{\psi}_{p_z E}(z) = -\tilde{\psi}_{p_z E}(z) + L_E \partial_E \tilde{\psi}_{p_z E}(z) = 0\,. \tag{28}$$

We could construct a solution for $\partial_E \tilde{\psi}_{p_z E}(z)$ by employing the Green's function, however, for our approximation we only need to consider,

$$\begin{aligned}
\int_{d/2}^{d/2} dz \tilde{\psi}^*_{p_z E}(z) L_E \partial_E \tilde{\psi}_{p_z E}(z) &= \int_{d/2}^{d/2} dz \tilde{\psi}^*_{p_z E}(z) \tilde{\psi}_{p_z E}(z) = dN^2_{p_z E} \\
&= -\tfrac{\hbar^2}{2m} \left( \begin{array}{l} \tilde{\psi}^*_{p_z E}(z) \partial_z \partial_E \tilde{\psi}_{p_z E}(z) \\ -(\partial_E \tilde{\psi}_{p_z E}(z)) \partial_z \tilde{\psi}^*_{p_z E}(z) \end{array} \right|_{-d/2}^{d/2} \\
&= -\tfrac{\hbar^2}{2m} [\tilde{\psi}^*_{p_z E}(d/2) \partial_E Z_{p_z E} - Z^*_{p_z E} \partial_E \tilde{\psi}_{p_z E}(d/2)] \\
&= -\tfrac{\hbar^2}{2m} \tilde{\psi}^*_{p_z E}(d/2) [\partial_E Z_{p_z E} - Z_{p_z E} \partial_E \ln \tilde{\psi}_{p_z E}(d/2)]
\end{aligned} \,. \tag{29}$$

Utilizing this result in (27), gives us,

$$\partial_E \left( \partial_z \ln \left| \tilde{\psi}_{p_z E}(z) \right|^2 \right|_{-d/2}^{d/2} = -\tfrac{2m}{\hbar^2} \frac{dN_{p_z E}}{\tfrac{1}{2} \left| \tilde{\psi}_{p_z E}(d/2) \right|^2}\,. \tag{30}$$

Expanding the denominator in the integral of (26), we have,

$$\partial_z \ln \left| \tilde{\psi}_{p_z E}(z) \right|^2 \Big|_{-d/2}^{d/2} \approx \partial_z \ln \left| \tilde{\psi}_{p_z E_{\cos\vartheta_0}}(z) \right|^2 \Big|_{-d/2}^{d/2} - \Delta E \, \tfrac{2m}{\hbar^2} \frac{dN^2_{p_z E_{\cos\vartheta_0}}}{\tfrac{1}{2} \left| \tilde{\psi}_{p_z E_{\cos\vartheta_0}}(d/2) \right|^2}\,, \tag{31}$$

and variation of the energy for the angular integration is,

$$\Delta E = -\tfrac{\hbar^2}{m} K_{\max} Q (\cos\vartheta - \cos\vartheta_0)\,, \tag{32}$$

Hence, our approximation of the angular integral over the singular factor is,

$$\tfrac{1}{2\pi} \int_{-\pi}^{\pi} d\vartheta \sin^2\vartheta \frac{1}{\tfrac{1}{2} \left( \partial_z \left| \tilde{\psi}_{p_z E^{\pm}_{n,\cos\vartheta}}(z) \right|^2 \right|_{-d/2}^{d/2}} \approx \frac{S(\alpha_{p_z E^{\pm}_{n,\cos\vartheta_0}} - \cos\vartheta_0 \mp i \,\mathrm{sgn}\, \varepsilon)}{2K_{\max} Q dN^2_{p_z E^{\pm}_{n,\cos\vartheta_0}}}\,. \tag{33}$$

The scaling parameter is,

$$\alpha_{p_z E_{n,\cos\vartheta_0}} = \left( \tfrac{1}{2} \partial_z \left| \tilde{\psi}_{p_z E}(z) \right|^2 \Big|_{-d/2}^{d/2} / dN^2_{p_z E} \right|_{E = E_{n,\cos\vartheta_0}} / \, 2K_{\max} Q\,, \tag{34}$$

and the energy is evaluated at,

$$E^{\pm}_{n,\cos\vartheta_0} = \xi_{k_z n} - \tfrac{\hbar^2}{2m} Q^2 \pm \hbar(\omega + i\varepsilon) - \tfrac{\hbar^2}{m} K_{\max} Q \cos\vartheta_0\,.$$

Combining (25) and (33), our approximation for angular integral of the Green's function is,

$$\begin{aligned}
\overline{G}_{p_z E_{n,\cos\vartheta_0}}(z, z') &= \tfrac{1}{2\pi} \int_{-\pi}^{\pi} d\vartheta \sin^2\vartheta \, G_{p_z E_{n,\cos\vartheta}}(z, z') \\
&\approx [g_{p_z E} G^{(1)}_{p_z E}(z, z') + \kappa^{-1}_{p_z E} \tilde{\psi}_{p_z E}(z) \tilde{\psi}^*_{p_z E}(z')]_{E = E_{n,\cos\vartheta_0}}
\end{aligned} \,, \tag{35a}$$

where,

$$\kappa^{-1}_{p_z E^{\pm}_{n,\cos\vartheta_0}} = \frac{S(\alpha_{p_z E^{\pm}_{n,\cos\vartheta_0}} - \cos\vartheta_0 \mp i \,\mathrm{sgn}\,\varepsilon)}{2K_{\max} Q d N^2_{p_z E^{\pm}_{n,\cos\vartheta_0}}} . \tag{35b}$$

The approximation must satisfy the periodic boundary conditions, and the boundary condition on the angular integral of the Green's function is clearly satisfied. In addition, we require,

$$\partial_z \overline{G}_E(d/2, z') - e^{ip_z d} \partial_z \overline{G}_E(-d/2, z') = \tfrac{2m}{\hbar^2}[-g_{p_z E} + \gamma_{p_z E} \kappa^{-1}_{p_z E}] / \tilde{\psi}^*_{p_z E}(d/2) = 0 , \tag{36}$$

and this requirement determines,

$$g_{p_z E} = \gamma_{p_z E} \kappa^{-1}_{p_z E} = \alpha_{p_z E} S(\alpha_{p_z E} - \cos\vartheta_0 \mp i \,\mathrm{sgn}\,\varepsilon) . \tag{37}$$

We employ an approximation where the wave functions are evaluated at energies within the angular integration where $\sin^2 \vartheta = 1$,

$$E^{\pm}_n = \xi_{k_z n} - \tfrac{\hbar^2}{2m} Q^2 \pm \hbar(\omega + i\varepsilon) , \tag{38}$$

and the coefficient is,

$$g_{p_z E^{\pm}_n} = \frac{\gamma_{p_z E^{\pm}_n}}{dN^2_{p_z E^{\pm}_n}} \frac{S(\alpha_{p_z E^{\pm}_n} \mp i \,\mathrm{sgn}\,\varepsilon)}{2K_{\max} Q} . \tag{39}$$

Our angular average of the Sternheimer function in (15) is expressed using these parameters.

It is instructive to consider the result we obtain when the energy coincides with the band energy,

$$E_{n,\cos\vartheta_{p_z n',n}} = \xi_{k_z n} - \tfrac{\hbar^2}{2m}[2K_{\max} Q \cos\vartheta_{p_z n',n} + Q^2] \pm \hbar\omega = \xi_{p_z n'} ,$$

and the coefficient, $\gamma_{p_z \xi_{p_z n'}} = 0$. We obtain,

$$\lim_{E \to \xi_{p_z n'}} \tfrac{1}{2\pi} \int_{-\pi}^{\pi} d\vartheta \sin^2 \vartheta \tilde{\psi}_{p_z E}(z) \tilde{\psi}^*_{p_z E}(z') \gamma^{-1}_{p_z E} \approx$$

$$\approx \frac{\tilde{\psi}_{p_z \xi_{p_z n'}}(z) \tilde{\psi}^*_{p_z \xi_{p_z n'}}(z')}{2K_{\max} Q d N^2_{\xi_{p_z n'}}} S(\cos\vartheta_{p_z n',n} \mp i \,\mathrm{sgn}\,\varepsilon) .$$

$$= \frac{\psi_{p_z n'}(z) \psi^*_{p_z n'}(z')}{2K_{\max} Q d} S(\cos\vartheta_{p_z n',n} \mp i \,\mathrm{sgn}\,\varepsilon)$$

This result mirrors the result, (8), we found before introducing the Sternheimer construction and it provides support for our approximations in evaluating the angular integral. However, it would approximate the angular integral with the result for a single band state.

**IV. Singularities of the Sternheimer Function**

The Sternheimer equation and its Green's function are singular when the energy coincides with the band energies [3]. Our approximation for the angular average Green's function treats these singularities in a consistent manner. In (18), we obtained,

$$\Delta^*_{p_z E,n} = \tfrac{\hbar^2}{2md} Z^*_{p_z E} \zeta_{p_z n}(d/2, E) ,$$

and, inserting this into (15), we have,

$$\bar{\zeta}_{p_z n}(z,E) \approx \frac{\tilde{\psi}_{p_z E}(d/2) Z^*_{p_z E}}{dN^2_{p_z E_0}} \frac{S(\alpha_{p_z E} \mp i\,\mathrm{sgn}\,\varepsilon)}{2K_{\max}Q} \zeta_{p_z n}(z,E)$$

$$= \frac{\tilde{\psi}_{p_z E}(d/2)\Delta^*_{p_z E,n}}{dN^2_{p_z E}} \frac{S(\alpha_{p_z E} \mp i\,\mathrm{sgn}\,\varepsilon)}{\frac{\hbar^2}{m}K_{\max}Q} \frac{\zeta_{p_z n}(z,E)}{\zeta_{p_z n}(d/2,E)} . \qquad (40)$$

Hence, we find that the integration over the singularities have been isolated and incorporated into $S(\alpha_{p_z E} \mp i\,\mathrm{sgn}\,\varepsilon)$,

$$\lim_{E\to\xi_{p_z n'}} \frac{\tilde{\psi}_{p_z E}(d/2)\Delta^*_{p_z E,n}}{dN^2_{p_z E}} S(\alpha_{p_z E} \mp i\,\mathrm{sgn}\,\varepsilon) = \psi_{p_z n'}(d/2)\Delta^*_{p_z n',n} S(\lim_{\alpha_{p_z E}\to 0} \alpha_{p_z E} \mp i\,\mathrm{sgn}\,\varepsilon)$$

In order to provide an assessment of our approximation, we consider the angular average of the Sternheimer function,

$$\tfrac{1}{2\pi}\int_{-\pi}^{\pi} d\vartheta \sin^2\vartheta\, \zeta_{p_z n}(z, E_{n,\cos\vartheta}) = \tfrac{1}{d}\sum_{n'} \psi_{p_z n'}(z)\Delta^*_{p_z n',n} S(\beta_{p_z n',E_n}) / \tfrac{\hbar^2}{m} K_{\max} Q , \qquad (41a)$$

where the parameter is,

$$\beta_{p_z n',E_n} = (\xi_{p_z n'} - E_n) / \tfrac{\hbar^2}{m} K_{\max} Q . \qquad (41b)$$

Our approximation for the angular average can also be expressed as a summation over all the band states. Using (9 and 15), we obtain,

$$\bar{\zeta}_{p_z n}(z,E_n) \approx \tfrac{1}{d}\sum_{n'} \frac{\psi_{p_z n'}(z)\Delta^*_{p_z n',n}}{\xi_{p_z n'} - E_n} \alpha_{p_z E_n} S(\alpha_{p_z E_n})$$

$$= \tfrac{1}{d}\sum_{n'} \frac{\psi_{p_z n'}(z)\Delta^*_{p_z n',n}}{\frac{\hbar^2}{m}K_{\max}Q} \alpha_{p_z E_n} S(\alpha_{p_z E_n}) / \beta_{p_z n',n} . \qquad (42)$$

The overlap of our wave function with that of a band state is obtained from,

$$\tfrac{1}{d}\int_{-d/2}^{d/2} dz \tilde{\psi}_{p_z E}(z)\, \mathrm{L}_E\, \psi^*_{p_z n'}(z) = (\xi_{p_z n'} - E)\tfrac{1}{d}\int_{-d/2}^{d/2} dz \psi^*_{p_z n'}(z)\tilde{\psi}_{p_z E}(z)$$

$$= \tfrac{\hbar^2}{2md}\psi^*_{p_z n'}(d/2) Z_{p_z E} .$$

Hence, we can express the scaling parameter, (16), as,

$$\alpha_{p_z E_n} = \frac{\gamma_{p_z E}}{2K_{\max}QdN^2_{p_z E}} = \frac{(\xi_{p_z n'} - E)}{\frac{\hbar^2}{m}K_{\max}QdN^2_{p_z E}} \frac{\tilde{\psi}^*_{p_z E}(d/2)}{\psi^*_{p_z n'}(d/2)} \int_{-d/2}^{d/2} dz \psi^*_{p_z n'}(z)\tilde{\psi}_{p_z E}(z)$$

$$= \beta_{p_z n',E} \frac{\tilde{\psi}^*_{p_z E}(d/2)}{\psi^*_{p_z n'}(d/2)} \frac{\int_{-d/2}^{d/2} dz \psi^*_{p_z n'}(z)\tilde{\psi}_{p_z E}(z)}{\int_{-d/2}^{d/2} dz \tilde{\psi}^*_{p_z E}(z)\tilde{\psi}_{p_z E}(z)} = \beta_{p_z n',E} \mathrm{I}_{p_z n',E} . \qquad (43)$$

Then we have,

$$\alpha_{p_z E} S(\alpha_{p_z E}) / \beta_{p_z n',E} = [1 + \sqrt{1 - (\beta_{p_z n',E}\mathrm{I}_{p_z n',E})^{-2}}\,]^{-1} \text{ for } |\alpha_{p_z E}| \ge 1, \qquad (44a)$$

and,

$$\alpha_{p_z E} S(\alpha_{p_z E}) / \beta_{p_z n',E} = \mathrm{I}^2_{p_z n',E}[\beta_{p_z n',E} \mp i\,\mathrm{sgn}\,\varepsilon \sqrt{\mathrm{I}^{-2}_{p_z n',E} - \beta^2_{p_z n',E}}\,] \text{ for } |\alpha_{p_z E}| \le 1 . \qquad (44b)$$

Hence, considering the limit at a singularity,

$$\lim_{E\to\xi_{p_z n'}} \mathrm{I}_{p_z n',E} = 1,$$

and, comparing  the exact and approximate expression, (41) and (42), we establish the agreement of our approximation of the angular average of the Sternheimer function with the exact result for the singular values of this function.

**V.  Concluding Remarks**

Our separation of the Green's function solution of the inhomogeneous differential equation satisfied by the Sternheimer function permits an approximation of the angular integral of the Green's function. The subsequent integration over the angular variable in the Sternheimer function resulted in an accurate treatment of the singularities in the direct evaluation of this function for the layered electron gas.

The poles in the susceptibility of the induced density response occur in the TDDFT calculations of many electron systems, and the Sternheimer procedure is often employed in the computations for these systems. In some of these systems it may prove useful to construct the Green's function and isolate its singular in order to provide an approximate which permits an accurate evaluation of the contributions from the singularities of the Sternheimer function.